\documentclass[aps,preprint,amsmath,amssymb,amsfonts,nofootinbib,superscriptaddress]{revtex4-1}
\usepackage{epsfig}
\usepackage{graphicx}
\usepackage{dcolumn}
\usepackage{bm}
\usepackage{amsthm}
\usepackage{amsmath}
\usepackage{color}
\usepackage{natbib} 
\usepackage[export]{adjustbox}
\usepackage{amsfonts}
\usepackage{amssymb}
\usepackage{braket}
\usepackage{xcolor}
\usepackage[hidelinks]{hyperref}
\hypersetup{
    colorlinks=false,
    linkcolor=false,
    filecolor=false,      
    urlcolor=false,
    pdfpagemode=FullScreen,
    }
\newcommand{\beq}{\begin{equation}}
\newcommand{\eeq}{\end{equation}}
\newcommand{\bea}{\begin{eqnarray}}
\newcommand{\eea}{\end{eqnarray}}

\begin{document}

\title{Odd Dimensional Nonlocal Liouville Conformal Field Theories}
\author{Amitay C. Kislev}
\affiliation{
 Raymond and Beverly Sackler School of Physics and Astronomy, Tel-Aviv University, Tel-Aviv 69978, Israel.
}
\author{Tom Levy}%
\affiliation{
 Raymond and Beverly Sackler School of Physics and Astronomy, Tel-Aviv University, Tel-Aviv 69978, Israel.
}
\author{Yaron Oz}%
\affiliation{
 Raymond and Beverly Sackler School of Physics and Astronomy, Tel-Aviv University, Tel-Aviv 69978, Israel.
}
\affiliation{Simons Center for Geometry and Physics, SUNY, Stony Brook, NY 11794, USA}
\date{\today}

\begin{abstract}
We construct Euclidean Liouville conformal field theories in odd number of dimensions.
The theories are nonlocal and non-unitary with 
a log-correlated Liouville field, a ${\cal Q}$-curvature background, and an exponential Liouville-type potential. 
We study the classical and quantum properties of these theories including the finite entanglement entropy part of the sphere partition function $F$, the boundary conformal anomaly and vertex operators' correlation functions.
We derive the analogue of the even-dimensional DOZZ formula and its semi-classical approximation.
\end{abstract}

\maketitle
\tableofcontents

\section{Introduction}

Since its discovery almost four decades ago \cite{Polyakov:1981rd}, two-dimensional Liouville conformal field theory (CFT) has become a cornerstone for diverse studies 
in field theory and string theory \cite{Seiberg:1990,Teschner:2001,Nakayama:2004}.
A higher-dimensional non-unitary CFT generalization of this theory emerged in the study of inertial range incompressible fluid turbulence as the field theory
of a Nambu-Goldstone dilaton mode \cite{Oz:2017ihc,Kolmogoro:194191,Oz:2018mdq}.
This higher-dimensional theory and its supersymmetric extentions  in even number of dimensions have been studied in
 \cite{Levy:2018bdc, Levy:2018xpu,Levy:2019tjl}.
 
The aim of this work is to study Liouville CFTs in odd number of dimensions, which  in contrast to the 
even-dimensional case, are nonlocal field theories. While  these theories are of interest by themselves, 
we note that the three-dimensional Liouville
CFT appears as part of the field theory of turbulence in three spatial dimensions  proposed in \cite{Oz:2017ihc,Oz:2018mdq}.
 
The action that defines Liouville CFT on a $d$-dimensional space ${\mathcal{M}_d}$, equipped with a metric $g_{ab}$, is formally the same in even and odd number of dimensions and reads:
\begin{equation}
S_L=\frac{d}{2 \Omega_{d}(d-1) !} \int_{\mathcal{M}_d} \sqrt{g} d^{d} x\left[\phi \mathcal{P}_{g} \phi+2 Q \mathcal{Q}_{g} \phi\right]+\mu \int_{\mathcal{M}_d} \sqrt{g} d^{d} x e^{d b \phi} \ ,
\label{action}
\end{equation}
where $\phi$ is the Liouville field and the dimensionless parameters are the background charge $Q$, the cosmological parameters $\mu$ and
$b$.  $\Omega_d = \frac{2 \pi^{\frac{d+1}{2}}}{\Gamma[\frac{d+1}{2}]}$ is the surface volume of the $d$-dimensional sphere. 
We will refer to the case $\mu=0$ as the Coulomb gas theory
$S_{CG} = S_L(\mu=0)$, and to the case $\mu=Q=0$ as the Gaussian theory, $S_{G} = S_L(\mu=0, Q=0)$.

In contrast to the even-dimensional case, the odd-dimensional action (\ref{action}) is nonlocal.
The conformally covariant GJMS operator  $\mathcal{P}_{g} $ \cite{GJMS}   is a linear self-adjoint pesudo-differential operator, whose principal part is
$(-\Delta)^{d / 2}$: 
\begin{equation}
\mathcal{P}_{g}= (-\Delta)^{d / 2}+\text {lower order terms} \ ,
\label{P}
\end{equation}
where $\Delta=g^{a b} \nabla_{a} \nabla_{b}$ is the Laplacian and $\nabla_{a}$ is the covariant derivative, $a=1,...,d$.
On $\mathbb{R}^d$ (or the unit ball $B^d$) it is a  fractional 
power of the Laplacian $(-\Delta)^{d / 2}$ and on $S^d$ it can be constructed via fractional powers of the conformal Laplacian. 
The $\mathcal{Q}$-curvature $\mathcal{Q}_{g}$  \cite{Q}  formally reads:
\begin{equation}
\mathcal{Q}_{g} = \frac{1}{2(d-1)} (-\Delta)^{d / 2-1}{\cal R} +\text {lower order terms}  \ ,
\label{Q}
 \end{equation}
where ${\cal R}$ is the scalar curvature, and we expect it to be nonlocal in general. 
Its integral over the manifold is an invariant of the conformal structure.

The stress-energy tensor $T_{ab}$ derived from (\ref{action}) is nonlocal, hence there are no local field theory charges generating the conformal symmetry.
As we will argue later in the article, we can formally express the A-type conformal anomaly on conformally flat spaces 
as \footnote{In three dimensions a necessary and sufficient condition for a manifold to be conformally flat is the vanishing of the Cotton tensor. In dimensions higher than three a necessary and sufficient condition is the vanishing of the Weyl tensor. Any simply connected compact conformally flat manifold is conformally equivalent to the round sphere.}:
\begin{equation}
\left<T^{a}_{a}\right> = a_d  {\mathcal{Q}}_{g}  \ , \label{T}
\end{equation}
where $a_d$ is the A-type conformal anomaly coefficient. In general, on the LHS of (\ref{T}) the stress-energy is nonlocal in the Liouville field and couples nonlocally to the background geometry, while on the RHS the ${\mathcal{Q}}_{g}$-curvature is nonlocal in the geometry variables. 

We consider in this work Liouville CFTs in $d=2n+1, n\geq 1$ dimensions.
The one-dimensional case $d=1$ is special since the
conformal transformations are equivalent to general coordinate transformations and it will be 
discussed separately. 
The outline of the paper is as follows. In section II
we study classical aspects of the Liouville CFT, including the Weyl symmetry, the classical field equations, the nonlocal GJMS operators and the 
${\mathcal{Q}}_{g}$-curvature. In section III we analyse the quantum properties of the theory. We calculate the boundary conformal anomaly, the entanglement entropy and the correlation functions of vertex operators and derive an analogue of the DOZZ formula. The details of the calculations are given in the appendices.

\section{Classical Nonlocal Liouville CFT}

\subsection{Weyl Symmetry}

Consider the Liouville CFT  defined by the action (\ref{action}). As in the even-dimensional case, in the odd-dimensional counterpart
under a  Weyl transformation of the metric:
\begin{equation}
g_{ab} \rightarrow e^{2 \sigma} g_{ab} \ ,
\label{weyl}
\end{equation}
the Liouville field $\phi$  transforms affinely as:
\begin{equation}
\label{eq:PhiTrans}
\phi \to \phi-Q \sigma \ ,
\end{equation}
while the conformal operator ${\cal P}_{g}$  transforms homogeneously as:
\begin{equation}
{\cal P}_{e^{2 \sigma} g} = e^{- d \sigma} {\cal P}_{g} \ , \label{Ptra}
\end{equation}
and the ${\cal Q}$-curvature  as:
\begin{equation}
{\cal{Q}}_{e^{2 \sigma} g} = e^{- d \sigma}\left({\cal{Q}}_{g} + {\cal P}_{g} \sigma \right) \ . \label{Qg}
\end{equation}
The action (\ref{action}) is then Weyl invariant:
\begin{equation}
\label{eq:ActionWeylTrans}
S_L(\phi-Q\sigma,e^{2\sigma}g) = S_L(\phi,g)-S_{CG}(Q\sigma,g) \ ,
\end{equation}
for the specific classical value of the background charge $Q = \frac{1}{b}$.

Equation (\ref{Qg}) can be written as:
\begin{equation}
{\cal P}_{g} \sigma + {\cal{Q}}_{g}  = {\cal{Q}}_{e^{2 \sigma} g} e^{d \sigma}  \ . \label{EMO}
\end{equation}
If we take ${\cal{Q}}_{e^{2 \sigma} g}$ to be a real constant  ${\cal Q}$ then solutions $\sigma$ to equation (\ref{EMO}) are  such that 
under a Weyl transformation $g_{ab} \rightarrow e^{2 \sigma} g_{ab}$ we get a conformally equivalent  manifold with
a constant ${\cal Q}$-curvature.

From equation (\ref{Qg}) follows that 
the integral of the ${\cal Q}$-curvature  on a closed Riemannian  manifold ${\cal M}_d$ is an invariant of the conformal structure:
\begin{equation}
\int_{{\mathcal{M}_d}}d^{d}x\sqrt{g}\mathcal{Q}_{g} = \frac{1}{2}\Omega_{d}(d-1)!\mathcal{Q}({\mathcal{M}_d}) \ . \label{E}
\end{equation}
When ${\cal M}_d$ is an even-dimensional conformally flat manifold, the ${\cal Q}$-curvature
is proportional to the Euler density and its integral is the Euler characteristic topological invariant, i.e.\ $\mathcal{Q}({\mathcal{M}_d}) = \chi( {\cal M} _d)$.
This is not the case
in odd number of dimensions. For instance,  while the Euler characteristic of odd-dimensional compact manifolds
vanishes, the  ${\cal Q}$-curvature does not. 
Specifically, we will see that $\mathcal{Q}(S^d)=2$.   %and 
%$\mathcal{Q}(B^d)=1$.

Keeping the normalization convention of \eqref{E}, we get that shifting the field by a constant affects the Coulomb gas action as:
\begin{align}
    S_{C G}\left[\phi+\phi_{0}\right] = \, S_{C G}[\phi]+\frac{\mathcal{Q}({\mathcal{M}_d})}{2}d Q\phi_{0} \ .
\end{align}

\subsection{Classical Field Equations}
By calculating the variation of the action \eqref{action} with respect to the Liouville field, we find the classical equations of motion:
\begin{equation}
    \label{eq:ClassicalEOM}
    \mathcal{P}_g \phi + Q\mathcal{Q}_g = -\Omega_d(d-1)!\mu b e^{db\phi} .
\end{equation}
For the classical value of the background charge as required by classical Weyl invariance, $Q = \frac{1}{b}$, this equation has a geometric interpretation as noted above. Equation \eqref{eq:PhiTrans} shows that we can think of the classical field $\phi_c = b\phi$ as a Weyl factor. Using the transformation law of the $\mathcal{Q}$-curvature \eqref{EMO}, we see that the EOM for $\phi_c$ is the equation defining a Weyl transformation from a metric $g_{ab}$ to a metric $e^{2\phi_c}g_{ab}$ that has a constant negative $\mathcal{Q}$-curvature:
\begin{equation}
    \mathcal{Q}_{e^{2\phi_c}g} = -\Omega_d(d-1)!\mu b^2 .
\end{equation}

\subsection{Spectral Decomposition}
\subsubsection{The Nonlocal GJMS Operator}

The spectral decomposition is a useful tool for studying nonlocal field theories. A natural basis in the Liouville case is the span of the eigenfunctions of $\mathcal{P}_{g}$.
For simplicity we consider only spaces where $\mathcal{P}_{g}$ can be written as a function of the Laplacian, i.e.\ $\mathcal{P}_{g} = f(-\Delta)$ where $f$ is a monotonic function on the set of positive real numbers. 
Two canonical examples for spaces for which this property holds are the sphere $S^d$ and the ball $B^d$. Using the \cite{Chang:1997} we have:
\begin{equation}
\begin{aligned}    
    \mathcal{P}_{g\left(B^d\right)} (-\Delta) = \left(-\Delta \right)^{\frac{d}{2}} , \quad 
    \mathcal{P}_{g\left(S^d\right)} (-\Delta) = \prod_{k=0}^{d-1} \left(-\Delta+\frac{k(d-k-1)}{R^2_S}\right)^{1/2} \ ,
\end{aligned}
\label{Laplacian}
\end{equation}
where $R_S$ is the radius of the sphere.
Denote by $V^{\nu}_p (x_a)$ the eigenfunctions of the Laplacian $\Delta$ with eigenvalue $-p^2$
and degeneracy index ${\nu}$, then
\begin{equation}
\mathcal{P}_g V^{\nu}_p (x_a) = f_{\mathcal{P}_g} (p^2) V^{\nu}_p (x_a) \ .
\end{equation}
The eigenfunctions and eigenvalues of the Laplacian on the sphere $S^d$ read:
\begin{equation}
    V_l^{\nu}(\vec{r}) = Y_{l,\nu}\left(\theta_{1}\ldots \theta_d\right), \quad p^2_l=\frac{1}{R_S^2}\, l(l+d-1) \ ,
\label{Sneigen}
\end{equation}
where $Y_{l,\nu}$ are the higher-dimensional spherical harmonics (see e.g.\ \cite{sh}). These satisfy $l\geq 0$, and $0 \leq \nu < K_{d,l}$, where $K_{d,l}$ is the degeneracy:
\begin{equation}
 K_{d,l} = \frac{2 l+d-1} {l} {l+d-2\choose l-1} \ . 
 \label{Kdl}
\end{equation}
The eigenfunctions of the Laplacian on the ball $B^d$ 
take the form \cite{Frye:2012jj}:
\begin{align}
\begin{aligned}
V^{J,\nu_1,\nu_2}_p (\vec{r}) =&  \, \frac{\mathcal{N}_{\nu_1}^{J} (p R_B)}{R_B} r^{1-d / 2} J_{\frac{d}{2}+\nu_1-1}(p r) Y_{\nu_1,\nu_2}\left(\theta_{1}\ldots \theta_{d-1}\right) , \\
V^{Y,\nu_1,\nu_2}_p (\vec{r}) =& \, \frac{\mathcal{N}_{\nu_1}^{Y} (p R_B)}{R_B} r^{1-d / 2} Y_{\frac{d}{2}+\nu_1-1}(p r) Y_{\nu_1,\nu_2}\left(\theta_{1}\ldots \theta_{d-1}\right) ,\\
\end{aligned}
\label{ef}
\end{align}
where $R_B$ is the radius of the ball and $J_{\nu}$, $Y_{\nu}$ are the Bessel functions of the first and second kind, respectively. In addition, we've defined the normalization constants:
\begin{equation}
\mathcal{N}_{\nu}^{f} (x) =  \left(\left(f_{\frac{d}{2}+\nu-1}(x)\right)^2-f_{\frac{d}{2}+\nu}(x)f_{\frac{d}{2}+\nu-2}(x)\right)^{-1/2}  \ ,
\end{equation}
where $f = Y, J$, which gives the normilzation constants for the first and second kind Bessel functions respectively.
The momenta are discretized,  $p_{\nu,n}=\frac{u_{\nu,n}}{R_B}$, where $u_{\nu,n}$ are the roots of the appropriate Bessel functions or their derivatives for Dirichlet or Neumann boundary conditions, respectively.
As a way to regularize the calculation in flat space $\mathbb{R}^d$, we preform calculations on the ball $B^d$ and take the limit of infinite radius, where equation \eqref{ef} remains valid, but with a different normalization as the momenta become continuous in the limit $p_{\nu, n} \to p$.

We will use the hemisphere $HS^{d}$ (given by $\theta_1\in [0, \pi/2]$) to calculate the boundary conformal anomaly.
The Laplacian eigenfunctions for the hemisphere are obtained from (\ref{Sneigen}) by taking the linear combinations $Y_{l,\nu}\left(\theta_{1}\ldots \theta_d\right) \rightarrow \frac{1}{2}( Y_{l,\nu}\left(\theta_{1}\ldots \theta_d\right) \pm Y_{l,\nu}\left(\pi-\theta_{1}\ldots \theta_d\right))$ with $-$ and $+$ corresponding to Dirichlet and Neumann boundary conditions, respectively. 
The corresponding eigenvalues and their degeneracies read:
\begin{equation}
p_l^2=\frac{l(l+d-1)}{R_S^2}, \qquad
K_{d,l,\text{Dirichlet}}={l+d-2\choose d-1}, \qquad
K_{d,l,\text{Neumann}}={l+d-1\choose d-1} \ . 
\label{deghem}
\end{equation}

\subsubsection{The $\mathcal{Q}$-Curvature of the Sphere}

One can calculate the $\mathcal{Q}$-curvature of conformally flat spaces $g_{ab}=e^{2 \sigma} \delta_{ab}$ \cite{Chang:1997}
by using the field equation: 
\begin{equation}
{\cal Q}_{e^{2 \sigma}\delta} = e^{-d \sigma} ({\cal P}_{\delta} \sigma + {\cal{Q}}_{\delta}) =  e^{-d \sigma} 
{\cal P}_{\delta} \sigma \ .
\end{equation}
By explicitly calculating the decomposition to $\mathcal{P}_g$ eigenfunctions, we can show that the stereographic projection, 
\begin{equation}
    \sigma = \log\left(\frac{2 R \rho}{\rho^2+r^2}\right) \ , \label{eq:SteographicWeylFactor}
\end{equation}
results in a space with a non-trivial constant $\mathcal{Q}$-curvature value, with $\rho, R$ being constant parameters with a dimension of length.  We use the decomposition in terms of the flat space eigenfunctions \eqref{ef}, and $\sigma$ being rotationally invariant and smooth implies that the relevant eigenfunctions are just $V^{J,0,0}_p (\vec{r}) =  \, \frac{r^{1-d / 2}}{\sqrt{\Omega_{d-1}}} \sqrt{p} J_{\frac{d}{2}-1}(p r)$. We get that the coefficients for this decomposition are given by:
\begin{eqnarray}
\sigma^{J,0,0}_p &\equiv& \int d^dr V^{J,0,0}_p (\vec{r}) \sigma = \sqrt{\Omega_{d-1}}  \int_0^\infty r^{d-1} dr\; r^{1-d / 2} \sqrt{p} J_{\frac{d}{2}-1}(p r) \log\left(\frac{2 R \rho}{\rho^2+r^2}\right)= \nonumber\\ 
&=& 2 \rho^{d/2} p^{-1/2} \sqrt{\Omega_{d-1}}K_{\frac{d}{2}} (\rho p)  \ ,
\end{eqnarray}
where we used the identity $\int_0^\infty dr r^{d/2} J_{\frac{d}{2}-1}(p r)  ({\rho^2+r^2})^{-\alpha}
= \frac{2^{1-\alpha}p^{-1} \rho^{d/2-\alpha}}{\Gamma(\alpha)} K_{\frac{d}{2}-\alpha} (\rho p)$, in the limit where $\alpha \to 0$ \cite{Bateman:1954}, and $K_{\nu}(x)$ is the modified Bessel function of the second kind. 
Therefore, when computing $\mathcal{P}_{\delta} \sigma$ we get:
\begin{equation}
\begin{aligned}
\mathcal{P}_\delta \sigma & = \int_0^\infty p^d \, dp \,\sigma^{J,0,0}_p V^{J,0,0}_p(r) \\
& = 2\rho^{\frac{d}{2}} r^{1-\frac{d}{2}} \int_0^\infty dp \, p^d K_{\frac{d}{2}} (\rho p) J_{\frac{d}{2}-1} (r p) = (d-1)! \frac{2^d\rho^d } {(\rho^2+r^2)^d} \ ,
\end{aligned}
\end{equation}
and we used $\int_{0}^{\infty}dp \, p^{\nu+\mu+1}J_\nu\left(py\right)K_\mu\left(pa\right)=\frac{2^{\nu+\mu}\Gamma\left(\nu+\mu+1\right)y^\nu a^\mu}{\left(a^2+y^2\right)^{\nu+\mu+1}}$.
 
 To compute the $\mathcal{Q}$-curvature of the unit sphere $S^d$, we note that the unit radius stereographic projection is exactly defined by the scale factor $\sigma(R=1) = \log\left(\frac{2\rho}{\rho^2+r^2}\right)$. From equation \eqref{EMO} we finally get the $\mathcal{Q}$-curvature,
\begin{align}
\mathcal{Q}_{g_{S^d}} = {\cal{Q}}_{e^{2 \sigma}\delta}= e^{-d \sigma} ({\cal P}_{\delta} \sigma + {\cal{Q}}_{\delta}) = (d-1)!\ .
\label{curvsphere}
\end{align}
Inserting this result into equation \eqref{E} gives us the integrated $\mathcal{Q}$-curvature of the sphere,
\begin{equation}
\mathcal{Q}\left(S^{d}\right) = \frac{2}{\Omega_{d}(d-1)!} \int_{S^{d}}d^{d}x\sqrt{g}\mathcal{Q}_{g_{S^d}} =  2 \ .
\end{equation}
The odd-dimensional conformal invariant has the same value as the Euler characteristic of the even-dimensional spheres.

\section{Quantum Nonlocal Liouville CFT}

In this section we study the quantum aspects of the odd-dimensional nonlocal Liouville CFT. We study the A-type conformal anomaly, find the finite part of the free energy in odd dimensions (entanglement entropy) and construct the correlation functions and structure constants of the theory.

\subsection{Conformal Anomaly}
We begin by studying the conformal anomaly of the free Coulomb gas theory $S_{CG}$. As for local CFTs, there is only a boundary quantum anomaly and no quantum bulk one. The bulk contribution to the trace anomaly is the classical contribution from the background charge.
The calculation of the anomaly follows 
\cite{Dowker:2010qy,Dowker:2013mba}.

Preforming the Gaussian integration of \eqref{action} for $\mu=0$, the quantum Coulomb gas action reads \cite{Levy:2018bdc}:
\begin{equation}
W = -\log(Z) = \frac{1}{2}\log\left(\det\left(\frac{d}{\Omega_{d}(d-1)!}\sqrt{g}\mathcal{P}_{g}\right)\right)-\frac{1}{2}\left(\frac{d}{\Omega_{d}(d-1)!}\right)^2Q^2\mathcal{Q}_{g}D\mathcal{Q}_{g} \ ,
\label{Wfirst}
\end{equation}
where $D$ is the propagator of the free theory:
\begin{equation}
(\sqrt{g}\mathcal{P}_{g})_x D(x,y) = \frac{\Omega_d (d-1)!}{d}\delta^d (x,y) \ ,
\label{eq:PropagatorDef}
\end{equation}
and we use the notation:
\begin{equation}
\mathcal{Q}_{g} D \mathcal{Q}_{g} \equiv \int d^{d}x d^{d}y \left(\sqrt{g}\mathcal{Q}_{g}\right)_x D(x,y)\left(\sqrt{g}\mathcal{Q}_{g}\right)_y \ .
\end{equation}
The second term in equation \eqref{Wfirst} is a completely geometric quantity which is basically a classical contribution to the effective action. It originates from completing the square in the Gaussian integration, and may be compared to the Weyl transformation law of the action \eqref{eq:ActionWeylTrans}. This equation shows that at the classical level the action shifts by a geometric quantity under Weyl transformation. The classical term in \eqref{Wfirst} is exactly responsible for maintaining this shift at the quantum level, i.e.\ the lack of Weyl invariance of this term has a classical origin.

We can explicitly calculate the geometric quantity $\mathcal{Q}_{g} D \mathcal{Q}_{g}$ for the sphere of radius $R$, whose metric is given by $g_{ab}=e^{2\sigma}\delta_{ab}$ with $\sigma$ as in equation \eqref{eq:SteographicWeylFactor}. According to \eqref{curvsphere} this space has $\mathcal{Q}_g = \mathcal{P}_{g}\sigma = (d-1)!R^{-d}$, and using \eqref{eq:PropagatorDef} we get:
 \begin{equation}
     \left(D \mathcal{Q}_{g}\right)_{S^{d}} \equiv \int d^{d}y D(x,y)\left(\sqrt{g}\mathcal{Q}_{g}\right)_y  = \int d^{d}y D(x,y)\left(\sqrt{g}\mathcal{P}_{g}\sigma\right)_y 
    = \frac{\Omega_d (d-1)!}{d}\sigma(x) .
 \end{equation}
We can now calculate $\mathcal{Q}_{g} D \mathcal{Q}_{g}$ for the sphere of radius $R$,
\begin{equation}
\begin{aligned}
    \left(\mathcal{Q}_{g} D \mathcal{Q}_{g}\right)_{S^{d}} &= \frac{\Omega_d (d-1)!}{d}\int d^{d}x \sqrt{g}\mathcal{Q}_{g}\sigma \\
    &= \frac{\Omega_d \Omega_{d-1} ((d-1)!)^2R^{-d}}{d}\int_{0}^{\infty} dr\,r^{d-1}\left(\frac{2\rho R}{\rho^2+r^2}\right)^d\log\left(\frac{2\rho R}{\rho^2+r^2}\right) \\
    &= \frac{(4\pi)^d \, \Gamma^2(d/2)}{d}\left(\log\left(\frac{R}{\rho}\right)+\log 2 +\psi^{(0)}\left(\frac{d}{2}\right)-\psi^{(0)}(d)\right) \ ,
    \label{eq:FreeEnQTerm}
\end{aligned}
\end{equation}
where $\psi^{(0)}$ is the polygamma function.

The conformal anomaly is given by the variation of the quantum action with respect to Weyl transformations,  
\begin{equation}
\label{eq:AnomalyDefinition}
\left<T^{a}_{a}\right>=\frac{1}{\sqrt{g}}\frac{\delta W}{\delta\sigma}  \ .
\end{equation}
On a conformally flat manifold, the requirement that the conformal anomaly \eqref{eq:AnomalyDefinition} must be a conformally covariant scalar function of the metric of conformal dimension $\Delta = d$ is a strict one. Since on these manifolds the metric is completely determined by a single scalar function, the Weyl factor, the only possibility for the conformal anomaly is to be proportional to the $\mathcal{Q}$-curvature, following equation \eqref{eq:AnomalyDefinition}, even in the odd-dimensional case.

The A-type anomaly coefficient (\ref{T}) is a sum of two terms 
\begin{equation}
a_d = A_d + B_d\, Q^2 \ ,  
\end{equation}
and from the functional derivative \eqref{eq:AnomalyDefinition} one gets 
\begin{equation}
B_d =  \frac{d}{\Omega_d (d-1)!}  \ .
\end{equation}
The anomaly part $A_d$ in odd number of dimensions has only a boundary contribution, which can be calculated as in \cite{Dowker:2010qy, Dowker:2013mba}. Consider the eigenvalues \eqref{deghem} of the GJMS operator \eqref{Laplacian} on
the odd-dimensional unit hemisphere $HS^d$:  
\begin{equation}
p_l^2 = \prod_{k=0}^{d-1}\sqrt{ \left( l+k\right)\left( l+d-1-k \right) } = \prod_{k=0}^{d-1}\left(\sum_{i=1}^d {m_i} +a+k \right) \ ,
\end{equation}
where the constant $a=\frac{1}{2}(d\pm 1)$ depends on the boundary conditions, and
$m_1,\dotsc,m_d$ are non-negative integers  that satisfy the constraint $l+\frac{d-1}{2}=\sum_{i=1}^d {m_i}+a$. In this representation, the degeneracies for the Dirichlet/Neumann boundary conditions \eqref{deghem} follow from the different choices of $m_1,\dotsc,m_d$ that satisfy the constraint:

\begin{equation}
\left( \sum_{\sum_{i} {m_i}=l+\frac{d-1}{2}-a} {1}\right)_{a=\frac{d-1}{2}} = {{l+d-1} \choose {d-1}}, \qquad  
\left( \sum_{\sum_{i} {m_i}=l+\frac{d-1}{2}-a} {1}\right)_{a=\frac{d+1}{2}} = {{l+d-2} \choose {d-1}} \ .
\label{Bterms}
\end{equation}

Using zeta function regularisation, we can write conformal anomaly (logarithmic) term in the expansion of the partition function \eqref{Wfirst} as:
\begin{equation}
\frac{1}{2}\lim_{s\rightarrow 0} \sum_{{m_1,\dotsc,m_d}} \frac{1}{\left[\prod_{k=0}^{d-1}\left(\sum_i {m_i} +a+k \right)\right]^s } \equiv \frac{1}{2}\lim_{s\rightarrow 0} Z_d (s,a)\ ,
\end{equation}
and evaluate it following \cite{Dowker:2010qy},
\begin{equation}
\begin{aligned}
    Z_d (0,a) =  \sum_{k=0}^{d-1} \frac{1}{d!} \zeta_d \left(0,a+k|\mathbf{1}_d\right) = & \frac{1}{d!} \sum_{k=0}^{d-1} \int_{0}^{1}dt\prod_{m=1}^{d}\left(a+k+t-m\right) \\ = &  \frac{1}{d!} \int_{0}^{d}dt\prod_{m=1}^{d}\left(t+a-\frac{d-1}{2}-m\right) \ ,
\end{aligned}
\label{zbar}
\end{equation}
where $\zeta_d$ is a Barnes zeta function.
Simplifying \eqref{zbar} we get 
$$ Z_d (0,a_D) = -Z_d (0,a_N) = \frac{1}{d!}  \int_{0}^{1}dt\prod_{m=1}^{d}\left(t-m\right) \ ,$$
where  $a_D=\frac{d-1}{2}$ and $a_N=\frac{d+1}{2}$ refer to Dirichlet and Neumann
boundary conditions, respectively. 

Thus, the anomaly coefficient $A_{d}$ takes the values,
\begin{equation}
A_d = \pm \frac{d}{2\Omega_d\left(d!\right)^2}\int_{0}^{1}dt\prod_{m=1}^{d}\left(t-m\right) \ ,
\end{equation}
where the positive and negative coefficients correspond to Dirichlet and Neumann boundary conditions, respectively.  Note, that summing the two boundary anomalies results in $A_d = 0$, which is the bulk 
sphere result.

\subsection{Entanglement Entropy}

The sphere partition function of a conformal field theory in odd number of dimensions, $d=2n+1$, generally has an expansion of the form:
\begin{equation}
    F_{S^{2n+1}} \equiv -\log Z_{S^{2n+1}} = c_{2n+1} (R \Lambda_{UV})^{2n+1} + 
    c_{2n-1} (R \Lambda_{UV})^{2n-1} + \cdot\cdot\cdot +c_{1} (R \Lambda_{UV}) + F_{2n+1} \ ,
    \label{W}
\end{equation}
where $R$ is the radius of the sphere and $\Lambda_{UV}$ is the UV cutoff.
The coefficients $c_k$ are regularization dependent, however, the universal finite number $F_{2n+1}$ is a physical observable
that can be interpreted as the entanglement entropy for a spherical entangling hypersurface.

In equations \eqref{Wfirst}, \eqref{eq:FreeEnQTerm} we saw that for the Liouville field theory, the sphere free energy also includes a logarithmic term in $R$, which is not accounted for in \eqref{W}. This term, together with an additional constant term, are proportional to the square of the background charge $Q^2$, and originate from a classical geometric term in the quantum action. The constant term proportional to $Q^2$ is ambiguous as a result of the classical logarithmic term, and its value is dependent on the non-physical re-scaling parameter $\rho$ in (\ref{eq:SteographicWeylFactor}). Thus, it cannot contribute to the physical observable $F_{2n+1}$, which is therefore independent of the background charge. The quantum contribution to $F_{2n+1}$ which is independent of $Q$ is well-defined and can be calculated.

In this section we will calculate finite part of the free energy $F_{2n+1}$ for the odd-dimensional Liouville conformal field theory.
Conformal invariance implies that $F_{2n+1}$ does not depend on marginal couplings of a conformal field theory \cite{Gerchkovitz:2014gta}. In the case of nonlocal Liouville theory that means that $F_{2n+1}$ does not depend on the cosmological constant $\mu$, and we can preform the calculation of the sphere partition function using the Gaussian theory.

The sphere partition function of the Gaussian theory is formally given by
\begin{equation}
    F = -\log\left(Z\right) = \frac{1}{2}\log\det\left(\mathcal{P}_{g\left(S^{d}\right)}\right) = \frac{1}{2}\sum_{l'=1}^{\infty} K_{d,l'}\log \Lambda_{l'} , 
    \label{eq:SphereF}
\end{equation}
where $\Lambda_{l'}$ are the eigenvalues of $\mathcal{P}_g$ on the sphere \eqref{Laplacian}, and $ K_{d,l'}$ are the corresponding degeneracies \eqref{Kdl}. In odd number of dimensions $d=2n+1$ the eigenvalues can be expressed as,
\begin{equation}
\label{eq:LambdaL}
    \Lambda_{l'} = \prod_{k=0}^{d-1}(l'+k) = (l'+n)\prod_{k=1}^n\left((l'+n)^2-k^2\right) = l\prod_{k=1}^n\left(l^2-k^2\right), 
\end{equation}
where we've defined $l=l'+n$. The degeneracies $K_{d,l'}$ can be recast as,
\begin{equation}
\label{eq:Degeneracies}
    K_{d,l'} = \frac{2 l'+d-1} {l'} {l'+d-2\choose l'-1} = \frac{2l(l+n-1)!} {(l-n)!(2n)!} = \frac{2}{(2n)!}l^2\prod_{k=1}^{n-1}\left(l^2-k^2\right) \ .
\end{equation}
The free energy is the formal infinite sum,
\begin{equation}
    F = \frac{1}{2}\sum_{l'=1}^{\infty}K_{d,l'}\log \Lambda_{l'} = \frac{1}{(2n)!}\sum_{l=n+1}^{\infty}l^2\prod_{k=1}^{n-1}\left(l^2-k^2\right)\log\left(l\prod_{k'=1}^{n}\left(l^2-k'^2\right)\right) .
    \label{eq:FreeEnergySum}
\end{equation}
Using the zeta function regularization we can extract the finite part of the sum $F_{2n+1}$. Using the regularization procedure detailed in appendix \ref{app:ZetaReg}, we get 
\begin{equation}
    \begin{aligned}
        F_{2n+1} &= \frac{n+\frac{1}{2}}{(2n)!}\sum_{j=1}^{n}(-1)^{j+1}\frac{(2j)!}{(2\pi)^{2j}}a_{j-1,n-1}\zeta(2j+1) - \left(n+\frac{1}{2}\right)\log n \\ 
        &\quad + \sum_{k=0}^{n}\int_{0}^{k}dt\left(\frac{\pi}{(2n)!}\prod_{j=0}^{n-1}\left(t^2-j^2\right)\cot(\pi t) + \frac{t}{n^2-t^2} \right) \ ,
        \label{eq:FinalFFormula}
    \end{aligned}
\end{equation}
where $a_{j,n}$ are defined as the coefficients of the polynomial $\prod_{k=1}^{n}\left[l^2-k^2\right] = \sum_{j=0}^n a_{j,n} l^{2j}$ (for more details see appendix A).

%An alternative derivation of the zeta regularization, provided in appendix \ref{app:MultiAnomaly}
Several values of $F_{2n+1}$ are listed in the following table as a function of the dimension:

\begin{center}
\begin{tabular}{ c | c }
 $d=2n+1$ & $F_{2n+1}$  \\ \hline
 $3$ & $\frac{3}{2}\frac{\zeta\left(3\right)}{\left(2\pi\right)^2}+\frac{1}{2}\log{\pi}$  \\  
 $5$ & $-\frac{5}{2}\frac{\zeta\left(5\right)}{\left(2\pi\right)^4}+\frac{55}{24}\frac{\zeta\left(3\right)}{\left(2\pi\right)^2}+\frac{1}{2}\log{\frac{\pi}{12}}$   \\
 $7$ & $\frac{7}{2}\frac{\zeta\left(7\right)}{\left(2\pi\right)^6}-\frac{77}{12}\frac{\zeta\left(5\right)}{\left(2\pi\right)^4}+\frac{133}{45}\frac{\zeta\left(3\right)}{\left(2\pi\right)^2}+\frac{1}{2}\log{\frac{\pi}{360}}$ \\
 $9$ & $-\frac{9}{2}\frac{\zeta\left(9\right)}{\left(2\pi\right)^8}+\frac{111}{8}\frac{\zeta\left(7\right)}{\left(2\pi\right)^6}-\frac{1781}{160}\frac{\zeta\left(5\right)}{\left(2\pi\right)^4}+\frac{11873}{3360}\frac{\zeta\left(3\right)}{\left(2\pi\right)^2}+\frac{1}{2}\log{\frac{\pi}{20160}}$ 
\end{tabular}
\end{center}

\subsection{Multiplicative Anomaly}
\label{SubSec:MultiplicativeAnomaly}
In the previous section we've calculated the functional determinant \eqref{eq:SphereF} of the nonlocal GJMS operator $\mathcal{P}_g$ operator on the sphere. To do so, we've written the corresponding zeta function and calculated its derivative at zero, as detailed in appendix \ref{app:ZetaReg}. 

As can be seen in \eqref{eq:LambdaL}, the eigenvalues on the sphere can be factored into a product of simple terms, each linear in the summation variable. If we could distribute the logarithm of the eigenvalues into a sum of logarithms of linear terms, then the calculation would be significantly simpler. However, such naive factorization is not always possible, since in general for two pseudo-differential operators $A, B$ we can have $\det(AB)\neq \det(A)\det(B)$. This is known as the multiplicative anomaly (for a review see e.g.\ \cite{Elizalde:1997nd}), and is a result of regularization.

Nevertheless, as it turns out in our specific case the multiplicative anomaly exactly vanishes, a fact which is proved in appendix \ref{app:MultiAnomaly}. This allows us to preform an alternative computation of the sphere partition function, which results in a different equivalent mathematical expression for the finite part of the free energy $F_{2n+1}$.

Knowing that there is no multiplicative anomaly, we can freely reorganize and manipulate the free energy sum,
\begin{equation}
    F = \frac{1}{2}\sum_{l'=1}^{\infty}K_{d,l'}\log\left(\prod_{k=0}^{d-1}\left(l'+k\right)\right) = \frac{1}{2}\sum_{k=0}^{d-1}\sum_{l'=1}^{\infty}K_{d,l'}\log\left(l'+k\right) \ .
    \label{eq:AltEnergySum}
\end{equation}
This sum can be further simplified by defining $l=l'+k$ and noting that for $-d+1<l'<0$ the formal expression for the degeneracy \eqref{Kdl} vanishes, i.e.\ $K_{d,l'}=0$, in addition to $K_{d,0}=1$. This allows us to extend the sum to $l'\geq-k+1 \implies l\geq1$, by subtracting the additional terms,
\begin{equation}
\begin{aligned}
    F &=\frac{1}{2} \sum_{l=1}^{\infty}  \left(\sum_{k=0}^{2n} K_{d,l-k} \right) \log(l) - \frac{1}{2} \sum_{k=1}^{2n} \sum_{l=1}^{k} K_{d,l-k} \log(l) \\
    &= \frac{1}{2} \sum_{l=1}^{\infty}  \left(\sum_{k=0}^{2n} K_{d,l-k} \right) \log(l) - \frac{1}{2} \log\left((2n)!\right)\ ,
\end{aligned}
\end{equation}
where again we've written the odd dimension as $d=2n+1$. We'll define the inner sum as a function of $l$, which by explicit calculation using \eqref{eq:Degeneracies} is given by:
\begin{equation}
 K_{n}(l) \equiv \sum_{k=0}^{2n} K_{d,l-k} =
\frac{2n+1+2l}{2n+1} {{2n+l} \choose {2n}} +\frac{2n+1-2l}{2n+1} {{l-1} \choose {2n}} \ .
\end{equation}
The function $K_{n}(l)$ is even with respect to $l$ so we can expand it as a polynomial with even powers only, $K_{n}(l) = \sum_{j=0}^{n}k_{j,n}l^{2j}$. Using $K_{n}(0) = 2$ we get:
\begin{equation}
    F= \frac{1}{2} \sum_{l=1}^{\infty} \sum_{j=1}^{n}{k_{j,n}l^{2j}}\log(l)+\sum_{l=1}^{\infty}\log(l)-\frac{1}{2}\log((2n)!) \ .
\end{equation}
We can now use zeta regularization, which gives us an alternative expression for $F_{2n+1}$, the finite part of the sphere partition function:
\begin{equation}
    F_{2n+1}  = \frac{1}{4}\sum_{j=1}^{n}{\left(-1\right)^{j+1}\frac{\left(2j\right)!}{\left(2\pi\right)^{2j}}k_{j,n}\zeta\left(2j+1\right)}+\frac{1}{2}\log{\frac{2\pi}{\left(2n\right)!}} \ .
    \label{eq:AltF}
\end{equation}

Here we've used zeta regularization to calculate the sum in the first line. We have verified that this new expression numerically agrees with \eqref{eq:FinalFFormula} for the values presented in the table, which demonstrates the lack of multiplicative anomaly.

\subsection{Correlation Functions}
We study the theory \eqref{action} on the sphere, and using the Weyl invariance of the action we preform a Weyl transformation to flat space. This transformation \eqref{eq:SteographicWeylFactor} is singular, and concentrates all of the non-trivial curvature at infinity. 

To regularize the action, we introduce an IR cutoff and study the theory on the ball $B^d$ with radius $R_B\to\infty$. The non-trivial curvature at infinity is then homogeneously spread on the boundary sphere $\partial B_d$. The action is defined as follows:
\begin{align}
    S_L = \frac{d}{2 \Omega_{d}(d-1) !} \int_{B_d} d^{d} x\, \phi (-\Delta)^\frac{d}{2}  \phi +\frac{d Q}{\Omega_{d-1}}\int_{\partial B_d}  d^{d-1} \Omega \, \phi + \mu \int_{B_d} d^{d} x e^{d b \phi} \ .
\label{actionsphere}
\end{align}

In this setting, the relevant eigenfunctions for the spectral decomposition are therefore the ball eigenfunctions \eqref{ef}.
The propagator in the Coulomb gas theory can be obtained with the help of IR regularization. Using the spectral decomposition for the nonlocal operator $(-\Delta)^\frac{d}{2}$, we get the following propagator, 
\begin{align}
    \langle\phi(x) \phi(y)\rangle = \mathcal{P}^{-1} \delta(x-y) \approx \frac{2}{d} \int_{\frac{\pi}{L_\infty}}^\infty  \frac{\cos{p|x-y|}}{p} dp = \frac{2}{d} \log \frac{L_\infty}{|x-y|}+O(1) ,
\label{propagator}
\end{align}
where $L_\infty$ is a the IR cutoff in length dimension.

Just as in the even dimensional theories, the vertex operators are defined as:
\begin{align}
    V_\alpha (x) = e^{d\alpha \phi(x)} \ .
    \label{eq:VertexOperator}
\end{align}
These are the primary conformal operators of the theory, with dimension given by:
\begin{align}
\Delta_{\alpha} = d \alpha(Q - \alpha).
\end{align}
Requiring that the potential term in the action $e^{db\phi}$ has the correct dimension, $\Delta_b=d$, determines the background charge:
\begin{align}
    Q = b+ \frac{1}{b} ,
\label{backgroundcharge}
\end{align}
and in the classic limit $b\rightarrow 0$ we go back to the classical value $ Q \to \frac{1}{b}$. 
In order to obtain the correlation functions of the C.G. theory, we use the 2-point function (\ref{propagator}), for vertex operators satisfying the condition of charge conservation $\sum_i \alpha _i = Q$:
\begin{align}
    \left\langle\prod_{i} V_{\alpha_{i}}\left(x_{i}\right)\right\rangle_{C.G.}=\prod_{i \neq i^{\prime}}\left|x_{i}-x_{i^{\prime}}\right|^{2 d \alpha_{i} \alpha_{i^{\prime}}} \ .
\end{align}
With the help of this result and by using the decomposition of the path integral measure into a zero mode $\phi_0$ and non-zero modes, i.e.\ $D \phi= D \phi\, d \phi_{0} \delta\left(\int_{\mathcal{M}} \sqrt{g} d^{d} x \phi -\phi_0\right)$, we can calculate a formal expression for the correlation functions in nonlocal Liouville theory, 
\begin{align}
\begin{aligned}
    \left\langle V_{\alpha_1}\left(x_{1}\right)\cdots  V_{\alpha_N}\left(x_{N}\right) \right\rangle & \equiv  \int D \phi e^{-S} \prod_{i=1}^{N} e^{d \alpha_{i} \phi(x_i)}
    \\ = \frac{\Gamma(-s) \mu^{s}}{d b} & \left\langle \left(\int d^{d}y e^{d b \phi\left(y\right)}\right)^s \times \prod_{i=1}^N e^{d \alpha_{i} \phi\left(x_{i}\right)}\right\rangle_{C G} ,
    \label{eq:CorrFuncDef}
\end{aligned}
\end{align}
where we have defined,
\begin{align}
s=\frac{Q-\sum_{i} \alpha_{i}}{b} ,
\end{align}
and $\left\langle \dots\right\rangle_{CG}$ refers to expectation value in the Coulomb Gas theory without zero-mode.
Thus, we see that the correlation function have poles at $s=n$ for non-negative integer $n$. We can write the residues at these poles as integrals over free field correlation functions:

\begin{align}
\begin{aligned}
\label{eq:residues}
    \underset{s=n}{\mathrm{Res}}\left\langle V_{\alpha_1}\left(x_{1}\right)\cdots  V_{\alpha_N}\left(x_{N}\right) \right\rangle & =
    \\ = \frac{(-\mu)^{n}}{n!} & \int d^{d}y_1\cdots d^{d}y_n \left\langle \prod_{i=1}^n V_{b}\left(y_{i}\right) \prod_{k=1}^N V_{\alpha_k}\left(x_{k}\right)\right\rangle_{C G}
    \\ =  \frac{(-\mu)^{n}}{n!} & \prod_{k>l = 1}^N \left|x_{k}-x_{l}\right|^{-2 d \alpha_{k} \alpha_{l}}
    \\  &\int d^{nd}y  \prod_{i>j=1}^n  \left|y_i-y_j\right|^{-2d b^{2}} \prod_{i=1}^n
    \prod_{k=1}^N \left|y_i-x_j\right|^{-2 db \alpha_{j} }  \ .
\end{aligned}
\end{align}
Note that the requirement $s=n$ can now be translated to a conservation of the background charge in the Coulomb Gas theory, i.e.\ $\sum_i\alpha_i + nb =  Q$.

Of particular interest in a CFT are the 3-point correlation functions, which by conformal invariance are constrained to have the form,
\begin{align}
    \label{eq:DefinitionSC}
    \left\langle V_{\alpha_1}\left(x_{1}\right)V_{\alpha_2} \left(x_{2}\right)V_{\alpha_3}\left(x_{3}\right) \right\rangle = \frac{C(\alpha_1,\alpha_2,\alpha_3)}{|x_1-x_2|^{\Delta_1+\Delta_2-\Delta_3} |x_2-x_3|^{\Delta_2+\Delta_3-\Delta_1} |x_3-x_1|^{\Delta_3+\Delta_1-\Delta_2}} . 
\end{align}
Here the coefficients $C(\alpha_1,\alpha_2,\alpha_3)$ are the structure constants of the theory, and they play a key role in determining the properties of the CFT.

In appendix \ref{app:StructConstsDerivation} we detail the calculation of the 3-point residue integrals following \cite{Furlan:2018jlv}, which also preformed the analytic continuation required to get the exact structure constants, 
\begin{align}
\begin{aligned}
    \label{eq:StructureConst}
    C\left(\alpha_1, \alpha_2, \alpha_3\right) = & \, 
    \left(-\frac{\pi^{d/2}}{\gamma_{\frac{d}{2}}\left(-\frac{d}{2}b^2\right)}\mu \left(\sqrt{\frac{d}{2}}b\right)^{-d(1+b^2)}\right)^{\frac{Q-\sum_i\alpha_i}{b}}  \frac{\sqrt{\frac{2}{d}}\Upsilon_{b}^{(d)'}(0)} {\Upsilon_{b}^{(d)}\left(\sum_i\alpha_i-Q\right)}
    \\ & \times \frac{\Upsilon_{b}^{(d)}(2\alpha_1)\Upsilon_{b}^{(d)}(2\alpha_2)\Upsilon_{b}^{(d)}(2\alpha_3)}
    {\Upsilon_{b}^{(d)}(\alpha_1+\alpha_2-\alpha_3)\Upsilon_{b}^{(d)}(\alpha_2+\alpha_3-\alpha_1)\Upsilon_{b}^{(d)}(\alpha_3+\alpha_1-\alpha_2)} \ .
\end{aligned}
\end{align}
As is further detailed in appendix \ref{app:StructConstsDerivation}, the special function $\Upsilon_b^{(d)}(x)$ is a $d$-dimensional generalization of the Upsilon function $\Upsilon_b(x)$, used in 2-dimensional Liouville CFT. To write its expression, we first define the special function $\Gamma_w(x)\equiv\frac{\Gamma_2(x|w,w^{-1})}{\Gamma_2(w/2+w^{-1}/2|w,w^{-1})}$, where $\Gamma_2(x|w_1, w_2)$ is the Double Gamma function. We can now define the $d$-dimensional generalization of the Upsilon function $\Upsilon_b$, namely
\begin{equation}
    \Upsilon_{b}^{(d)}(x) \equiv \frac{1}{\Gamma_{\sqrt{\frac{d}{2}}b}\left(\sqrt{\frac{d}{2}}x\right)\Gamma_{\sqrt{\frac{d}{2}}b}\left(\sqrt{\frac{d}{2}}(Q-x)\right)} \ . 
\end{equation}
This definition satisfies $\Upsilon_{b}^{(d)}(x)=\Upsilon_{b}^{(d)}(Q-x)$, and in the two-dimensional case we return to the usual Upsilon function, $\Upsilon_b^{(2)}(x)=\Upsilon_b(x)$.

\subsection{Semi-Classical Approximation}
The semi-classical limit of Liouville CFT is given by $b \rightarrow 0$. In this limit the background charge (\ref{backgroundcharge}) approaches its classical value and the path integral defining the correlation functions \eqref{eq:CorrFuncDef} is dominated by the saddle point of the action. In this limit we can use the saddle point approximation to calculate the semi-classical value of the correlators. 
In order to simplify the calculations, we define the classical Liouville field $\phi_c = b \phi$.
The action on the sphere \eqref{actionsphere} can then be written in this limit as:
\begin{align}
\begin{aligned}
S_L (\phi_c) = \frac{d}{2\Omega_d (d-1)! b^2} \int_{B^d} d^d x \, \phi_c \mathcal{P}_g \phi_c+\mu\int_{B^d} e^{d\phi_c} +\frac{d}{\Omega_{d-1}b^2} \int_{\partial B^d} d^{d-1}\Omega \phi_c+O(b^0).
\end{aligned}
\end{align}

In this section we will calculate the semi-classical limit of correlation functions of "light" vertex operators \eqref{eq:VertexOperator}, defined as operators $V_{\alpha_{i}}$ with $\alpha_i=O(b)$. As we have shown in the even dimensional case \cite{Levy:2018bdc}, in these conditions the corresponding path integral has no real saddle points. To overcome this, we consider correlation functions with fixed area $A$, which is defined as $A=\int d^dx e^{db\phi}$, by inserting into the path integral the identity element $1=\int_{0}^{\infty}dA\delta\left(A-\int d^dx e^{db\phi}\right)$. This insertion gives us:
\begin{equation}
\begin{aligned}
 \left\langle V_{\alpha_1}\left(x_{1}\right)\cdots  V_{\alpha_N}\left(x_{N}\right) \right\rangle &= \int \frac{dA}{A} e^{-\mu A} \left\langle V_{\alpha_1}\left(x_{1}\right)\cdots  V_{\alpha_N}\left(x_{N}\right) \right\rangle_{A} \\
 &= (\mu A)^s \Gamma(-s) \left\langle V_{\alpha_1}\left(x_{1}\right)\cdots  V_{\alpha_N}\left(x_{N}\right) \right\rangle_{A} \ .
\end{aligned}
\end{equation}
Here $\left\langle V_{\alpha_1}\left(x_{1}\right)\cdots  V_{\alpha_N}\left(x_{N}\right) \right\rangle_{A}$ is the fixed area correlation function, and we have used equation \eqref{eq:CorrFuncDef}. The fixed area path integral for light vertex operators does have real saddle points, which allows us to calculate the structure constants \eqref{eq:DefinitionSC} in the semi-classical approximation. The saddle points are given by solutions to the fixed area equation of motion, which can be obtained by using a Lagrange multiplier:
\begin{align}
(-\Delta)^{d/2}\phi_c = \frac{\Omega_d (d-1)!}{A} e^{d\phi_c} .
\end{align}
Thus saddle points correspond to Weyl factors that describe manifolds with positive constant $\mathcal{Q}$-curvature and finite area. As we have shown in equations \eqref{eq:SteographicWeylFactor}, \eqref{curvsphere}, one such solution is the Weyl factor of the sphere with radius $A$:
\begin{align}
\phi_c  = -\log\left(\frac{\rho^2+r^2}{\rho}\right)+ \frac{1}{d} \log\left(\frac{2^d A}{\Omega_d}\right) \ .
\label{eq:SphereASolution}
\end{align}
Following the calculation done previously in \cite{Levy:2018bdc}, by the conformal invariance of the theory, all other relevant saddle points are given by conformal transformations of the solution \eqref{eq:SphereASolution}. The semi-classical approximation of the correlation functions is then given by summing over all saddle points, i.e.\ integrating over the moduli space of saddle points, which is the conformal group modulo its subgroup that leaves \eqref{eq:SphereASolution} invariant. When changing the path integral into an integral over this moduli space we introduce some $b$-dependent Jacobian $\mathcal{A}(b)$, which can be shown to satisfy $\log(\mathcal{A}(b))=O(\log b)$. Finally, explicitly preforming the integration for the 3-point function of light vertex operators, we get the semi-classical approximation of the structure constants  for $\alpha_i=O(b)$:
\begin{align}
\begin{aligned}
C(\alpha_1, \alpha_2, \alpha_3)  \approx & \frac{\pi^{\frac{d}{2}}}{2}  \mathcal{A}(b) \Gamma\left(\frac{\sum_i\alpha_i-Q}{b}\right)\left(\frac{\Omega_d }{2^d}\mu\right)^{\frac{Q-\sum_i\alpha_i}{b}}e^{-\frac{S_{bulk}}{b^2}}\Gamma\left[\frac{d}{2b}(\alpha_1+\alpha_2+\alpha_3)\right] \\
& \times \frac{\Gamma\left(\frac{d}{2b}(\alpha_2+\alpha_3-\alpha_1)\right)\Gamma\left(\frac{d}{2b}(\alpha_3+\alpha_1-\alpha_2)\right)\Gamma\left(\frac{d}{2b}(\alpha_1+\alpha_2-\alpha_3)\right)}{\Gamma\left(\frac{d}{b}\alpha_1\right)\Gamma\left(\frac{d}{b}\alpha_2\right)\Gamma\left(\frac{d}{b}\alpha_3\right)} \ .
\label{eq:SemiClassicalStruct}
\end{aligned}
\end{align}
Here $S_{bulk}$ is the value of the bulk action defined by,
\begin{equation}
    S_{bulk} = \frac{d}{2\Omega_d(d-1)!} \int_{B^d} d^d r \phi_c (-\Delta)^{d/2} \phi_c,
\end{equation}
which is evaluated for the solution \eqref{eq:SphereASolution} for the unit sphere, i.e. $A=\Omega_d$. Preforming this calculation gives us the value:

\begin{equation}
S_{bulk} = \frac{2^{d-1} d}{\Omega_d } \int d^dr \left( \frac{-\log(1+r^2)}{(1+r^2)^d} \right) =  - \frac{d}{2} \left(\psi^{(0)} (d)-\psi^{(0) } \left(\frac{d}{2}\right)\right)\ .
\label{SA}
\end{equation}

\section{Discussion}

Liouville conformal field theories provide examples of solvable interacting CFTs in dimensions higher that two. In this work we considered these theories in odd number of dimensions where they are nonlocal.
We studied their classical and quantum properties and found that they exhibit similar structures to that of the local even-dimensional Liouville CFTs.

There are various directions to follow.
We derived the exact DOZZ formula \eqref{eq:StructureConst},
and independently its semi-classical approximation (\ref{eq:SemiClassicalStruct}).
Matching the sum over saddle points calculation and the limit $b\rightarrow 0$ of the exact calculation is an interesting open problem that requires a generalization of the methods used for the even-dimensional case in \cite{Furlan:2018jlv}.

The stress energy tensor of these theories is nonlocal. It will be important to construct it,
understand the structure of the nonlocal generators of the conformal algebra and calculate the charges such as $C_T$ of these theories. In fact, a general classification of possible anomaly terms of these nonlocal field theories is still lacking.

Renormalization group flow properties of higher-dimensional Liouvllile theories are largely unexplored.
Being non-unitary in dimensions higher than two, it is not clear whether there are monotonicity theorems
corresponding to anomaly and $F$ charges.
Finally, adding boundaries and defects to these theories may shed new light on the structure of these theories.

\section*{Acknowledgements}

The work is supported in part by  Israel Science Foundation Center of Excellence.

\appendix

\section{Zeta Regularization of the Sphere Partition Function}
\label{app:ZetaReg}
In this section we compute the finite part of the formal infinite sum \eqref{eq:FreeEnergySum},
\begin{equation}
  F = \frac{1}{2}\sum_{l=l_0}^{\infty}K_{d,l}\log\left(l\prod_{k=1}^{n}\left(l^2-k^2\right)\right) \ ,  
\end{equation} by employing the technique of zeta regularization. We begin by writing the identity,
\begin{equation}
    F = -\frac{1}{(2n)!}\left.\frac{d}{ds}\right|_{s=0}\sum_{l=n+1}^{\infty}l^2\prod_{k=1}^{n-1}\left(l^2-k^2\right)\left(l\prod_{k'=1}^{n}\left(l^2-k'^2\right)\right)^{-s} .
    \label{eq:FullFZetaFunc}
\end{equation}
Written this way, we note that for $s<1$ the expression we sum over actually vanishes for $l=0,1,\dots,n-1$. Therefore, we can replace the sum over $l\geq n+1$ with a sum over $l\neq n$.

We can then expand the eigenvalues \eqref{eq:LambdaL} and degeneracies \eqref{eq:Degeneracies} as a polynomial in $l$ using,
\begin{equation}
    \prod_{k=1}^{n}\left[l^2-k^2\right] = \sum_{j=0}^n a_{j,n} l^{2j},\quad a_{j,n} =   \left(-1\right)^{j+n}  \sum_{1\le k_1<k_2\ldots<k_j\le n} \frac{\left[n!\right]^2}{\prod_{i=1}^{j}k_i^2} \ ,
\label{appendixpoly}\end{equation}
and use this expression to simplify $F$. We express each term in the sum over eigenvalues as:
\begin{equation}
\begin{aligned}
\label{eq:LambdaMultinomExp}
    \Lambda_l^{-s} &= l^{-s}\left(\sum_{j=0}^n a_{j,n}  l^{2j}\right)^{-s}  \\
    &= l^{-(2n+1)s}\sum_{M = 0}^{\infty}\sum_{m_0+\dots+m_{n-1} = M}{-s \choose M}{M \choose m_0,\dots,m_{n-1}}\prod_{j=0}^{n-1}\left(a_{j,n} l^{-2(n-j)}\right)^{m_j} \\
    &= \sum_{M = 0}^{\infty}\sum_{\sum_j m_j = M}{-s \choose M}{M \choose m_0,\dots,m_{n-1}} \left(\prod_{j=0}^{n-1}a_{j,n}^{m_j}\right) l^{-(2n+1)s-2\sum_{j} (n-j)m_j} ,
\end{aligned}
\end{equation}
where we  used $a_{n,n} = 1$ and the generalized multinomial theorem, with multinomial coefficients defined as:
\begin{equation}
    {M \choose m_0,\dots,m_{n-1}} \equiv \frac{M!}{m_0!m_1!\cdots m_{n-1}!} . 
\end{equation}
Inserting this into the free energy, and denoting $m \equiv \sum_{j=0}^{n-1} (n-j)m_j$, we get:
\begin{equation}
\begin{aligned}
    F &= -\frac{1}{(2n)!}\left.\frac{d}{ds}\right|_{s=0}\sum_{l=0, l\neq n}^{\infty}\sum_{i=0}^{n-1}a_{i,n-1}l^{2(i+1)}\Lambda_l^{-s} \\
    &= -\frac{1}{(2n)!}\left.\frac{d}{ds}\right|_{s=0}\sum_{i=0}^{n-1}\sum_{M\geq 0, \sum_j m_j = M}{-s \choose M}{M \choose \{m_j\}} \left(\prod_{j=0}^{n-1}a_{j,n}^{m_j}\right)a_{i,n-1}\sum_{l=0, l\neq n}^{\infty}l^{-(2n+1)s-2m+2(i+1)} .
\end{aligned}
\end{equation}
Using the identity ${-s \choose M} = \delta_{M,0}+(-1)^{M}\frac{1}{M}s+O(s^2)$, we can compute the derivative at $s=0$,
\begin{equation}
\begin{aligned}
    F &= -\frac{2n+1}{(2n)!}\sum_{i=0}^{n-1} a_{i,n-1}\left(\zeta'(-2(i+1)) + n^{2(i+1)}\log n\right) + \\
    &\quad -\frac{1}{(2n)!}\sum_{i=0}^{n-1}\sum_{M\geq 1, \sum_j m_j = M}\frac{(-1)^M}{M} {M \choose \{m_j\}} \left(\prod_{j=0}^{n-1}a_{j,n}^{m_j}\right)a_{i,n-1}\left(\zeta\left(2(m-i-1)\right) - n^{-2(m-i-1)}\right).
\label{Fterm}
\end{aligned}
\end{equation}
While the second line seems complicated, it can be made significantly simpler by noting some basic identities. Looking at equation \eqref{eq:LambdaMultinomExp}, one can write,
\begin{equation}
\label{eq:LambdaDerivativeExp}
    -\left.\frac{d}{ds}\right|_{s=0} \Lambda_{l}^{-s} = (2n+1)\log l - \sum_{M=1}^{\infty}\sum_{\sum_j m_j = M}\frac{(-1)^M}{M} {M \choose \{m_j\}} \left(\prod_{j=0}^{n-1}a_{j,n}^{m_j}\right)l^{-2m} .
\end{equation}
On the other hand, using $\log \Lambda_l = -\left.\frac{d}{ds}\right|_{s=0} \Lambda_{l}^{-s}$, we have
\begin{equation}
\begin{aligned}
\label{eq:LambdaLogTaylor}
    \log \Lambda_l &= \log\left(l\prod_{k=1}^n\left(l^2-k^2\right)\right) \\
    & = (2n+1)\log l - \sum_{k=1}^{n}\log\left(1-k^2l^{-2}\right) \\
    &= (2n+1)\log l - \sum_{m=1}^{\infty} \frac{\sum_{k=1}^{n} k^{2m}}{m}l^{-2m} .
\end{aligned}
\end{equation}
Expressions \eqref{eq:LambdaDerivativeExp} and \eqref{eq:LambdaLogTaylor} are both analytic functions in $l^{-1}$ (discarding the identical log term) since in these calculations we kept $l$ arbitrary. Thus, the coefficients of the two power series must match, and we get
\begin{equation}
     \sum_{M=1}^{\infty}\sum_{\sum_j m_j = M}\frac{(-1)^M}{M} {M \choose \{m_j\}} \left(\prod_{j=0}^{n-1}a_{j,n}^{m_j}\right) = \frac{\sum_{k=1}^{n} k^{2m}}{m} .
\end{equation}
Applying this identity to (\ref{Fterm}), we get for the finite part:
\begin{equation}
\label{eq:FreeEnDoubleSum}
\begin{aligned}
    F_{2n+1} &= -\frac{2n+1}{(2n)!}\sum_{i=0}^{n-1} a_{i,n-1}\zeta'(-2(i+1)) -\left(n+\frac{1}{2}\right) \log(n)\\
    &\quad -\sum_{k=1}^{n}\sum_{m=1}^{\infty}\frac{k^{2m}
    }{m}\left(\frac{1}{(2n)!}\sum_{i=0}^{n-1}a_{i,n-1}\zeta\left(2(m-i-1)\right) -\frac{1}{2} n^{-2m}\right) ,
\end{aligned}
\end{equation}
where we used $\sum_{i=0}^{n-1}a_{i,n-1}n^{2(i+1)}=n^2\prod_{k=1}^{n-1}\left(n^2-k^2\right) = \frac{1}{2}(2n)!$.

Equation \eqref{eq:FreeEnDoubleSum} demonstrates that even when using zeta regularization the order of summation can be rearranged such that: 
\begin{equation}
    \sum_{l=l_0}^{\infty}K_{d,l}\log\left(l\prod_{k=1}^{n}\left(l^2-k^2\right)\right) = 
    \sum_{l=l_0}^{\infty}K_{d,l}\log(l)+\sum_{k=1}^{n}\sum_{l=l_0}^{\infty}K_{d,l}\log\left(l^2-k^2\right) \ .
    \label{eq:NoMultiAnom1}
\end{equation}
As explained in subsection \ref{SubSec:MultiplicativeAnomaly}, this is a highly non-trivial result, which shows the multiplicative anomaly vanishes for this decomposition of the eigenvalues into simpler products. In appendix \ref{SubSec:MultiplicativeAnomaly} we prove that we can further decompose into liner terms with no multiplicative anomaly.

To simplify \eqref{eq:FreeEnDoubleSum}, we will use the following values of the derivative of the zeta function,
\begin{equation}
    \zeta'(-2j) = (-1)^j\frac{(2j)!}{2(2\pi)^{2j}}\zeta(2j+1), \quad j\in \mathbb{N} \ .
\end{equation}
Note, that we have the following generating function for the positive even integer values of the zeta function,
\begin{equation}
    -\frac{\pi}{2}t\cot(\pi t) = \sum_{j=0}^{\infty}\zeta(2j)t^{2j} ,
\end{equation}
while for negative even integer we have $\zeta(-2j) = 0,\; j\in \mathbb{N}$. We can now write an integral expression for the infinite sum remaining in $F_{2n+1}$ for each $k$,
\begin{equation}
\begin{aligned}
    \sum_{m=1}^{\infty}\frac{k^{2m} 
    }{m} & \left(\frac{1}{2} n^{-2m} - \frac{1}{(2n)!}\sum_{i=0}^{n-1}a_{i,n-1}\zeta\left(2(m-i-1)\right)\right) \\
    & = \int_{0}^{k}dt \left(\sum_{m=1}^{\infty}t^{2m-1}n^{-2m}-\frac{2}{2n!}\sum_{m=1}^{\infty}\sum_{i=0}^{n-1}a_{i,n-1}t^{2m-1}\zeta\left(2(m-i-1)\right)\right) \\
    & = \int_{0}^{k}dt \left(\frac{t}{n^2-t^2}-\frac{2}{(2n)!}\sum_{i=0}^{n-1}a_{i,n-1}t^{2i+1}\sum_{m=0}^{\infty}\zeta(2(m-i))t^{2(m-i)}\right) \\
    & = \int_{0}^{k}dt \left(\frac{t}{n^2-t^2}+\frac{\pi}{(2n)!}\prod_{j=0}^{n-1}\left(t^2-j^2\right)\cot(\pi t) \right) \ .
    \label{eq:ShiftedLogZeta}
\end{aligned}
\end{equation}
Combing this result with equation \eqref{eq:FreeEnDoubleSum} gives us our final result \eqref{eq:FinalFFormula}.

We note that the zeta regularization technique employed here agrees with previous results when applied to simpler operators. Specifically, we have checked that when applying this exact method to the conformal Laplacian, we get the same results for the finite part of the free energy as in \cite{Klebanov:2011gs}.

\section{No Multiplicative Anomaly for the GJMS Operator}
\label{app:MultiAnomaly}

%{\color{red} Edit}
In this appendix we show that the multiplicative anomaly vanishes when expanding the logarithm of the eigenvalues in the infinite sum \eqref{eq:SphereF} into logarithms of linear terms. During our derivation in appendix \ref{app:ZetaReg}, we have shown that multiplicative anomaly vanishes for the partial decomposition \eqref{eq:NoMultiAnom1}. What remains to be shown is that this infinite sum can be further decomposed by writing the terms inside the logarithm as $l^2-k^2=(l+k)(l-k)$, with no multiplicative anomaly. 

The degeneracy $K_{d,l}$ is an even polynomial in $l$, so in fact we can prove a stronger statement by replacing it with a general even power of $l$, i.e.\ $l^{2N}$ for a non-negative integer $N$. We define the  quantities:
\begin{equation}
F_\pm=\sum_{l=k+1}^{\infty}{l^{2N}\log{\left(l\pm k\right)}}, \quad F_0=\sum_{l=k+1}^{\infty}{l^{2N}\log{\left(l^2-k^2\right)}}, \quad \Delta_k=F_0-F_+-F_- \ .
\end{equation}
To show that the multiplicative anomaly vanishes we need to prove $\Delta_k=0$ and we shall do that by induction on $k$. Trivially the base case holds, $\Delta_0 = 0$.  Shifting the summation index in $F_{\pm}$ we get:
\begin{align}
\begin{aligned}
    F_{+} + F_{-} &= \sum_{l=1}^{\infty}(l-k)^{2N}\log(l) + \sum_{l=1}^{\infty}(l+k)^{2N}\log(l) - \sum_{l=-k+1}^{k}l^{2N}\log(l+k) = \\
    &= -2\sum_{j=0}^{N}{2N \choose 2j}k^{2(N-j)}\zeta'(-2j)- \sum_{l=1}^{k}l^{2N}\log(l+k)- \sum_{l=1}^{k-1}l^{2N}\log(k-l) \ ,
\end{aligned}
\end{align}
where we used $\sum_{l=1}^{\infty}l^{\alpha}\log(l) = -\left.\frac{d}{ds}\right|_{s=0}\sum_{l=1}^{\infty}l^{\alpha-s}=-\zeta'(-\alpha)$. Similarly, we can express $F_0$ as follows,
\begin{align}
\begin{aligned}
    F_0 &= \sum_{l=k+1}^{\infty} l^{2N}\log\left(l^2\right) + \sum_{l=k+1}^{\infty}{l^{2N}\log{\left(1-k^2l^{-2}\right)}} \\
    &= -2\zeta'(-2N) - 2\sum_{l=1}^{k}l^{2N}\log(l)+\sum_{l=k+1}^{\infty}{l^{2N}\log{\left(1-k^2l^{-2}\right)}} \ .
\end{aligned}
\end{align}
Using a very similar derivation as in  \eqref{eq:ShiftedLogZeta}, we can compute the remaining infinite sum,
\begin{equation}
\sum_{l=k+1}^{\infty}{l^{2N}\log{\left(1-k^2 l^{-2}\right)}}
= \int_{0}^{k}dt\left(\pi t^{2N}\cot{\left(\pi t\right)}+\sum_{l=1}^{k}\frac{l^{2N}}{l-t}\right)-\sum_{l=1}^{k}l^{2N}\log\left(\frac{l+k}{l}\right) \ .
\end{equation}
Note, that the integrand is finite in the integration interval, as the poles of the two terms cancel each other. We can simplify the expression by explicitly integrating the second term and obtain for $F_0$:
\begin{equation}
\begin{aligned}
F_0 = - & 2\zeta^\prime\left(-2N\right)+\int_{0}^{k}dt\left(\pi t^{2N}\cot{\left(\pi t\right)}+\frac{k^{2N}}{k-t}\right)-k^{2N}\log 2-\sum_{l=1}^{k-1}l^{2N}\log\left(k^2-l^2\right) \ .
\end{aligned}
\end{equation}
This integration step introduces poles into the integration interval, so when applicable the integral should be understood as the Cauchy principal value.

The multiplicative anomaly, $\Delta_k=F_0-F_+-F_-$, reads
\begin{equation}
\Delta_k = 2\sum_{j=0}^{N-1}{2N \choose 2j}k^{2(N-j)}\zeta'(-2j)+\int_{0}^{k}dt\left(\pi t^{2N}\cot{\left(\pi t\right)}+\frac{k^{2N}}{k-t}\right)+k^{2N}\log(k) \ .
\end{equation}

In order to prove that $\Delta_{k}=0$ for any $k$ we assume that it vanishes for a particular value of $k$ and consider the difference 
\begin{equation}
\begin{aligned}
\Delta_{k+1}-\Delta_{k} = & \, 2\sum_{j=0}^{N-1}\sum_{p=2j+1}^{2N}{k^{2N-p}\frac{(2N)!}{(2j)!(p-2j)!(2N-p)!}}\zeta^\prime\left(-2j\right) \\ 
& +\int_{k}^{k+1}dt\left(\pi t^{2N}\cot{\left(\pi t\right)}+\frac{\left(k+1\right)^{2N}-k^{2N}}{k+1-t} \right) \ .
\end{aligned}
\end{equation}
Note that since $\Delta_k=0$, by our assumption, we have $\Delta_{k+1}-\Delta_k=\Delta_{k+1}$. The first term can be recast, using the re-summation $\sum_{j=0}^{N-1}\sum_{p=2j+1}^{2N} ...=\sum_{p=1}^{2N}\sum_{j=0}^{{\left\lfloor (p-1) /2\right\rfloor}} ...$. 

Furthermore, the integral can be simplified by the integration parameter change $u=t-k$, which finally gives the result,
\begin{equation}
\begin{aligned}
\Delta_{k+1}=\sum_{p=1}^{2N}{{2N\choose p}}k^{2N-p}\left(2\sum_{j=0}^{\left\lfloor\frac{p-1}{2}\right\rfloor}{{p \choose 2j}}\zeta^\prime\left(-2j\right)  +\int_{0}^{1}du\left(\pi u^p\cot{\left(\pi u\right)}+\frac{1}{1-u}\right) \right) \ .
\end{aligned}
\end{equation}
The sum of two terms in the parentheses vanishes for each value of $p$ by an integral identity, leaving us with $\Delta_k = 0$ for all non-negative integer $k$ by induction. This completes the proof that the multiplicative anomaly vanishes for this decomposition.

\section{Exact Correlation Functions by Relation to a Free Field}
\label{app:StructConstsDerivation}
In order to calculate the exact correlation function we first need to calculate it's residues using the relation to a free field \eqref{eq:residues}. To solve the relevant conformal integrals, we can use a mathematical integral identity of Liouville CFT. This section is a review of the work done in \cite{Furlan:2018jlv}. 

The starting point for solving these integrals is utilising the reflection identity:
\begin{equation}
    V_{Q-\alpha} = R(\alpha)V_{\alpha} , 
\end{equation}
where the function $R(\alpha)$ is called the reflection coefficient. We can then look at the possible residues of the correlation function \eqref{eq:residues} and the same correlation function with all operators $V_{\alpha}$ replaced by their reflected operators $V_{Q-\alpha}$. If a pole exists in both versions then we can require:
\begin{equation}
    \underset{Q - \sum_i \alpha_i=nb}{\mathrm{Res}}\left\langle V_{\alpha_1}\left(x_{1}\right)\cdots  V_{\alpha_N}\left(x_{N}\right) \right\rangle = \prod_{i=1}^N R(\alpha_i)^{-1} \underset{Q - \sum_i (Q-\alpha_i)=mb}{\mathrm{Res}}\left\langle V_{Q-\alpha_1}\left(x_{1}\right)\cdots  V_{Q-\alpha_N}\left(x_{N}\right) \right\rangle , 
\end{equation}
for some non-negative integers $n,m$. This in turn implies the conformal integral identity:
\begin{align}
\begin{aligned}
\label{eq:IntegralDual}
 &\frac{(-\mu)^n}{n!} \prod_{k>l = 1}^N \left|x_{k}-x_{l}\right|^{-2 d \alpha_{k} \alpha_{l}} \int d^{nd}y  \prod_{i>j=1}^n  \left|y_i-y_j\right|^{-2d b^{2}} \prod_{i=1}^n
    \prod_{k=1}^N \left|y_i-x_k\right|^{-2 db \alpha_{k} }= \\
    &\frac{(-\mu)^m}{m!}\prod_{i=1}^N R(\alpha_i)^{-1}  \prod_{k>l = 1}^N \left|x_{k}-x_{l}\right|^{-2 d (Q-\alpha_{k}) (Q-\alpha_{l})} \int d^{md}y  \prod_{i>j=1}^m  \left|y_i-y_j\right|^{-2d b^{2}} \prod_{i=1}^m
    \prod_{k=1}^N \left|y_i-x_k\right|^{-2 db (Q-\alpha_{k}) } .
\end{aligned}
\end{align}
For the two residues to exist, we have the background charge conservation requirements:
\begin{equation}
    Q-\sum_{i=1}^{N} \alpha_i=nb,
    \quad Q- \sum_{i=1}^N (Q-\alpha_i)=mb \ .
\end{equation}
Summing these two equations we get a relation which determines the number of external points $N$ in the correlation function:
\begin{equation}
    N = -\frac{b}{Q}(n+m)+2 .
\end{equation}
Since $n, m$ can be any non-negative integers, then for $N$ to be a positive integer $(-Q/b)$ must be a non-negative integer. To fulfil this requirement we are going to analytically continue the Liouville CFT to a specific imaginary value of $b$. This procedure will in turn provide us with a conformal integral identity which can be used for the general $b$ case.

To get an identity for all possible $N$-point function,  i.e.\ all integer $N\leq 2$, we must choose $(-Q/b)=1$, which using $Q=1/b+b$ implies:
\begin{equation}
    Q = -b, \quad b^2=-\frac{1}{2},\quad N = n+m+2 .
\end{equation}
Inserting these specific values into the analtically continued equation \eqref{eq:IntegralDual} and taking the limit $x_N\to \infty$
\begin{align}
\begin{aligned}
 &\frac{(-\mu)^n}{n!} \int d^{nd}y  \prod_{i>j=1}^n  \left|y_i-y_j\right|^{d} \prod_{i=1}^n
    \prod_{k=1}^{n+m+1} \left|y_i-x_k\right|^{-2 db \alpha_{k} }=\frac{(-\mu)^m}{m!}\prod_{i=1}^{n+m+1} R(\alpha_i)^{-1}R(\alpha_N)^{-1}\times \\
    &  \prod_{k>l = 1}^{n+m+1} \left|x_{k}-x_{l}\right|^{-2 d (Q^2-2Q(\alpha_{k}+\alpha_l))} \int d^{md}y  \prod_{i>j=1}^m  \left|y_i-y_j\right|^{d} \prod_{i=1}^m
    \prod_{k=1}^{n+m+1} \left|y_i-x_k\right|^{-2 db (Q-\alpha_{k}) } .
\end{aligned}
\end{align}
Finally, lets define $D_s(y) \equiv \prod_{i>j=1}^s  \left|y_i-y_j\right|$, $a_i \equiv db\alpha_i$ and $r(a_i) = \left.R(\alpha_i)^{-1}\right|_{b^2=-1/2}$. Using charge conservation we have $\alpha_N = Q-nb-\sum_{i=1}^{N-1}\alpha_i \implies a_N = d(n+1)/2-\sum_{i=1}^{n+m+1}a_i$ , and we can rewrite the integral identity as:
\begin{align}
\begin{aligned}
 &\frac{(-\mu)^n}{n!} \int d^{nd}y  D_n^{d}(y) \prod_{i=1}^n
    \prod_{k=1}^{n+m+1} \left|y_i-x_k\right|^{-2 a_k }=\frac{(-\mu)^m}{m!}\prod_{i=1}^{n+m+1}r(a_i)r\left(d(n+1)/2-\textstyle\sum_{i=1}^{n+m+1}a_i\right)\times \\
    &  \prod_{k<l}^{n+m+1}|x_k-x_l|^{d-2a_k-2a_l} \int d^{md}y  D_m^{d}(y) \prod_{i=1}^m
    \prod_{k=1}^{n+m+1} \left|y_i-x_k\right|^{-d+2a_k } .
\end{aligned}
\end{align}
The only thing left to obtain the full identity is to find the function $r(a_i)$, which can be found by looking at the 3-point function $n=1, m=0, N=3$:
\begin{align}
\begin{aligned}
 \int d^{d}y \left|y-x_1\right|^{-2 a_1 }\left|y-x_2\right|^{-2 a_2 }&=(-\mu)^{-1}r(a_1)r(a_2)r\left(d-a_1-a_2\right) \left|x_1-x_2\right|^{d-2a_1-2a_2} \\
 & = \pi^{d/2}\frac{\Gamma\left(\frac{d}{2}-a_1\right)\Gamma\left(\frac{d}{2}-a_2\right)\Gamma\left(a_1+a_2-\frac{d}{2}\right)}{\Gamma\left(a_1\right)\Gamma\left(a_2\right)\Gamma\left(d-a_1-a_2\right)} \ .
\end{aligned}
\end{align}
Choosing the value $\mu = -\pi^{-d/2}$ we get the reflection coefficient:
\begin{equation}
    r(a_i) = \frac{\Gamma\left(\frac{d}{2}-a_i\right)}{\Gamma(a_i)} \equiv \frac{1}{\gamma_{\frac{d}{2}}(a_i)} = \gamma_{\frac{d}{2}}(d/2-a_i) .
\end{equation}
Finally, the full conformal integral identity is:
\begin{align}
\begin{aligned}
\label{eq:IntIdentity}
 & \int dy_n  D_n^{d}(y) \prod_{i=1}^n
    \prod_{k=1}^{n+m+1} \left|y_i-x_k\right|^{-2 a_k }=\prod_{i=1}^{n+m+1}\frac{1}{\gamma_{\frac{d}{2}}(a_i)}\frac{1}{\gamma_{\frac{d}{2}}\left(d(n+1)/2-\textstyle\sum_{i=1}^{n+m+1}a_i\right)}\times \\
    &  \prod_{k<l}^{n+m+1}|x_k-x_l|^{d-2a_k-2a_l} \int dy_m  D_m^{d}(y) \prod_{i=1}^m
    \prod_{k=1}^{n+m+1} \left|y_i-x_k\right|^{-d+2a_k } ,
\end{aligned}
\end{align}
where for simplicity we've defined the measure $dx_n \equiv d^{nd}x/\left(\pi^{nd/2} n!\right)$.
To return to the general residue integral \eqref{eq:residues}, which includes the factor $D_{n}^{-2db^2}(y)$, we will look at the identity \eqref{eq:IntIdentity} with $n\to n-1, m = 0$ and $a_i = d(1+b^2)/2$. We get:
\begin{align}
\begin{aligned}
\label{eq:IntId1}
D_{n}^{-2db^2}(y) &= D_{n}^{d}(y)D_{n}^{-d-2db^2}(y) \\
&= D_{n}^{d}(y) \frac{\gamma_{\frac{d}{2}}\left(-n\frac{d}{2}b^2\right)}{\gamma_{\frac{d}{2}}\left(-\frac{d}{2}b^2\right)^n}\int du_{n-1}  D_{n-1}^{d}(u) \prod_{i=1}^{n-1} \prod_{k=1}^{n} \left|y_i-u_k\right|^{-d(1+b^2) } .
\end{aligned}
\end{align}
The last identity we need is the result of \eqref{eq:IntIdentity} for $m=0$, $a_i=d(1+b^2)/2$ for $i=3,\dots,n-1$, which we rewrite as:
\begin{align}
\begin{aligned}
\label{eq:IntId2}
 & \int dy_n  D_n^{d}(y) \prod_{k=1}^n
    \prod_{i=1}^{n-1} \left|y_i-u_k\right|^{-d(1+b^2)}\left|y_i-x_1\right|^{-2a_1}\left|y_i-x_2\right|^{-2a_2}\\
    &  = \frac{\gamma_{\frac{d}{2}}\left(-\frac{d}{2}b^2\right)^{n-1}  D_{n-1}^{-d-2db^2}(u)\left|x_{12}\right|^{d-2a_1-2a_2}}{\gamma_{\frac{d}{2}}(a_1)\gamma_{\frac{d}{2}}(a_2)\gamma_{\frac{d}{2}}\left(d-a_1-a_2-(n-1)\frac{d}{2}b^2\right)}\prod_{i=1}^{n-1}\left|u_i-x_1\right|^{-2a_1-db^2 }\left|u_i-x_2\right|^{-2a_1-db^2 } .
\end{aligned}
\end{align}
Going back to the original integral for the 3-point function \eqref{eq:residues}, with $x_3 \to \infty$, we have:
\begin{align}
\begin{aligned}
    I_n(a_1, a_2, a_3) \equiv & \int dy_n  D_n^{-2db^2}(y) \prod_{i=1}^n \left|y_i-x_1\right|^{-2a_1}\left|y_i-x_2\right|^{-2a_2} \\
    = & \, \frac{\gamma_{\frac{d}{2}}\left(-n\frac{d}{2}b^2\right)}{\gamma_{\frac{d}{2}}\left(-\frac{d}{2}b^2\right)^n}\int du_{n-1}  D_{n-1}^{d}(u)dy_n  D_n^{d}(y) \prod_{k=1}^n \prod_{i=1}^{n-1} \left|y_i-u_k\right|^{-d(1+b^2)}\left|y_i-x_1\right|^{-2a_1}\left|y_i-x_2\right|^{-2a_2} \\
    = & \, \frac{\gamma_{\frac{d}{2}}\left(-n\frac{d}{2}b^2\right)\left|x_{12}\right|^{d-2a_1-2a_2}}{\gamma_{\frac{d}{2}}\left(-\frac{d}{2}b^2\right)\gamma_{\frac{d}{2}}(a_1)\gamma_{\frac{d}{2}}(a_2)\gamma_{\frac{d}{2}}\left(a_3+(n-1)\frac{d}{2}b^2\right)} \\
    & \times \int du_{n-1}  D_{n-1}^{-2db^2} (u) \prod_{i=1}^{n-1}\left|u_i-x_1\right|^{-2a_1-db^2}\left|u_i-x_2\right|^{-2a_1-db^2 } \\
    =& \, \frac{\gamma_{\frac{d}{2}}\left(-n\frac{d}{2}b^2\right)\left|x_{12}\right|^{d-2a_1-2a_2}}{\gamma_{\frac{d}{2}}\left(-\frac{d}{2}b^2\right)\gamma_{\frac{d}{2}}(a_1)\gamma_{\frac{d}{2}}(a_2)\gamma_{\frac{d}{2}}\left(a_3+(n-1)\frac{d}{2}b^2\right)} I_{n}\left(a_1+\frac{d}{2}b^2, a_2+\frac{d}{2}b^2, a_3\right) ,
\end{aligned}
\end{align}
where we've applied \eqref{eq:IntId1} to get from the first equality to the second, and equation \eqref{eq:IntId2} to get from the second to the third. We've also used the condition $a_3 = d-(n-1)db^2-a_1-a_2$. This equation translates to a recursion relation on the residues of the structure constants:
\begin{align}
\begin{aligned}
    C_n(\alpha_1, \alpha_2, \alpha_3) =& \frac{\gamma_{\frac{d}{2}}\left(-n\frac{d}{2}b^2\right)}{\gamma_{\frac{d}{2}}\left(-\frac{d}{2}b^2\right)\gamma_{\frac{d}{2}}(db \alpha_1)\gamma_{\frac{d}{2}}(db\alpha_2)\gamma_{\frac{d}{2}}\left(db\alpha_3+(n-1)\frac{d}{2}b^2\right)} C_{n-1}\left(\alpha_1+\frac{b}{2}, \alpha_2+\frac{b}{2}, \alpha_3\right) \\
    = & \left(\frac{-\pi^{d/2}\mu}{\gamma_{\frac{d}{2}}\left(-\frac{d}{2}b^2\right)}\right)^n\prod_{k=0}^{n-1}\frac{\gamma_{\frac{d}{2}}\left((k-n)\frac{d}{2}b^2\right)}{\gamma_{\frac{d}{2}}\left(db\alpha_1+k\frac{d}{2}b^2\right)\gamma_{\frac{d}{2}}\left(db\alpha_2+k\frac{d}{2}b^2\right)\gamma_{\frac{d}{2}}\left(db\alpha_3+k\frac{d}{2}b^2\right)} .
\end{aligned}
\end{align}

Given the residues, we can now preform analytic continuation to find the general expression for the structure constants. To do so, we use the special function $\Gamma_w(x)\equiv\frac{\Gamma_2(x|w,w^{-1})}{\Gamma_2(w/2+w^{-1}/2|w,w^{-1})}$, where $\Gamma_2(x|w_1, w_2)$ is the Double Gamma function. The function $\Gamma_w(x)=\Gamma_{w^{-1}}(x)$ then satisfies
\begin{equation}
    \label{eq:DoubleGamma}
    \frac{\Gamma_{w}(x+w)}{\Gamma_{w}(x)} = \sqrt{2\pi}\frac{w^{wx-\frac{1}{2}}}{\Gamma(wx)} .
\end{equation}
We can now define the $d$-dimensional generalization of the Upsilon function $\Upsilon_b$, namely
\begin{equation}
    \label{eq:UpsilonDef}
    \Upsilon_{b}^{(d)}(x) \equiv \frac{1}{\Gamma_{\sqrt{\frac{d}{2}}b}\left(\sqrt{\frac{d}{2}}x\right)\Gamma_{\sqrt{\frac{d}{2}}b}\left(\sqrt{\frac{d}{2}}(Q-x)\right)} \ . 
\end{equation}
 Following \eqref{eq:DoubleGamma}, this function then satisfies the relation $\Upsilon_{b}^{(d)}(x) = \Upsilon_{b}^{(d)}(Q-x)$ and the functional relation

\begin{equation}
    \frac{\Upsilon_{b}^{(d)}(x+b)}{\Upsilon_{b}^{(d)}(x)} = \left(\sqrt{\frac{d}{2}}b\right)^{d/2-dbx}\gamma_{\frac{d}{2}}\left(\frac{d}{2}bx\right) .
\end{equation}

% When shifting the argument by $\frac{1}{b}$ we get an additional functional relation,

% \begin{equation}
%     \frac{\Upsilon_{b}^{(d)}\left(x+\frac{1}{b}\right)}{\Upsilon_{b}^{(d)}(x)} = \left(\sqrt{\frac{d}{2}}b\right)^{d\frac{x}{b}+\frac{(d-2)}{b^2}-\frac{d}{2}}\prod_{k=0}^{\frac{d}{2}-1}\gamma_1\left(\frac{x}{b}+\frac{2k}{db^2}\right) \ .
%     \label{eq:UpsilonFucntionalRel2}
% \end{equation}

We can then write the following products,
\begin{equation}
    \prod_{k=0}^{n-1}\frac{1}{\gamma_{\frac{d}{2}}\left(db\alpha+k\frac{d}{2}b^2\right)} = \left(\sqrt{\frac{d}{2}}b\right)^{\frac{d}{2}nb\left(Q-4\alpha-nb\right)} \frac{\Upsilon_{b}^{(d)}(2\alpha)}{\Upsilon_{b}^{(d)}(2\alpha+nb)} \ .
\end{equation}
Using the recursion relation, we can also compute the residues of $\Upsilon_{b}^{(d)}(x)$
\begin{equation}
    \underset{x=-nb}{\mathrm{Res}} \frac{\Upsilon_{b}^{(d)}(b)}{\Upsilon_{b}^{(d)}(x)} = \frac{\left(\sqrt{\frac{d}{2}b}\right)^{d/2-1}}{\Gamma(d/2)}\left(\sqrt{\frac{d}{2}}b\right)^{\frac{d}{2}nb(Q+nb)}\prod_{k=0}^{n-1}\gamma_{\frac{d}{2}}\left((k-n)\frac{d}{2}b^2\right) \ .
\end{equation}
Finally, preforming the analytic continuation by replacing $n$ with the continuous variable $s = \frac{Q-\sum_i \alpha_i}{b}$ we get the structure constants \eqref{eq:StructureConst}.

\end{document}